\documentclass[aps,prl,twocolumn,amsmath,amssymb,nofootinbib,superscriptaddress,floatfix,reprint,shortbibliography]{revtex4-2}
\usepackage[dvips]{graphicx}
\usepackage{latexsym}
\usepackage{amsmath}
\usepackage{amsfonts}
\usepackage{amssymb}
\usepackage{bm}
\usepackage{color}
\usepackage{txfonts}
\usepackage{float}
\usepackage{url}
\usepackage[colorlinks=true, urlcolor=blue, linkcolor=blue, citecolor=blue, pdftex]{hyperref}
\usepackage{ulem}
\usepackage{graphicx}
\usepackage{extpfeil}
\usepackage{subfigure}
\usepackage{physics}
\usepackage{setspace}
\DeclareUnicodeCharacter{2212}{-}

\begin{document}

\newcommand{\fig}[2]{\includegraphics[width=#1]{#2}}
\newcommand{\la}{{\langle}}
\newcommand{\ra}{{\rangle}}
\newcommand{\dg}{{\dagger}}
\newcommand{\upa}{{\uparrow}}
\newcommand{\dna}{{\downarrow}}
\newcommand{\ab}{{\alpha\beta}}
\newcommand{\ias}{{i\alpha\sigma}}
\newcommand{\ibs}{{i\beta\sigma}}
\newcommand{\hH}{\hat{H}}
\newcommand{\hn}{\hat{n}}
\newcommand{\hc}{{\hat{\chi}}}
\newcommand{\hU}{{\hat{U}}}
\newcommand{\hV}{{\hat{V}}}
\newcommand{\br}{{\bm r}}
\newcommand{\bk}{{{\bm k}}}
\newcommand{\bq}{{{\bm q}}}
\newcommand{\bp}{{{\bm p}}}
\def\gsim{~\rlap{$>$}{\lower 1.0ex\hbox{$\sim$}}}
\setlength{\unitlength}{1mm}
\newcommand{{\vhf}}{$\chi^\text{v}_f$}
\newcommand{{\vhd}}{$\chi^\text{v}_d$}
\newcommand{{\vpd}}{$\Delta^\text{v}_d$}
\newcommand{{\ved}}{$\epsilon^\text{v}_d$}
\newcommand{{\vved}}{$\varepsilon^\text{v}_d$}
\newcommand{\pprl}{Phys. Rev. Lett. \ }
\newcommand{\pprb}{Phys. Rev. {B}}
\newcommand{\LNO}{La$_3$Ni$_2$O$_7$ }

 \title{Jahn–Teller distortion on strained \LNO thin films}
 \author{Yuxin Wang}
\affiliation{Kavli Institute for Theoretical Sciences, University of Chinese Academy of Sciences,
	Beijing, 100190, China}
 
 \author{Zhan Wang}
 \email{zhan.wang@iphy.ac.cn}
 \affiliation{Beijing National Laboratory for Condensed Matter Physics and Institute of Physics, Chinese Academy of Sciences, Beijing 100190, China}

\author{Fu-Chun Zhang}
\email{fuchun@ucas.ac.cn}
\affiliation{Kavli Institute for Theoretical Sciences, University of Chinese Academy of Sciences,
	Beijing, 100190, China}

 \author{Kun Jiang}
 \email{jiangkun@iphy.ac.cn}
 \affiliation{Beijing National Laboratory for Condensed Matter Physics and Institute of Physics, Chinese Academy of Sciences, Beijing 100190, China}
 \affiliation{School of Physical Sciences, University of Chinese Academy of Sciences, Beijing 100190, China}

\date{\today}

\begin{abstract}
    We present a systematic study of the electronic structure of strained La$_3$Ni$_2$O$_7$ thin films. We show that biaxial compressive strain mainly elongates the outer apical Ni-O bond while leaving the inner apical Ni-O bond nearly unchanged. As a result, the Jahn--Teller splitting $\Delta_{JT}$ is strongly enhanced, whereas the interlayer $d_{z^2}$ hopping $t_\perp^z$ changes only weakly. Since superconductivity is widely believed to emerge only below a critical in-plane lattice constant, our results identify the strain-enhanced $\Delta_{JT}$ as the relevant microscopic tuning parameter. Consistently, the calculated Fermi surfaces and Hall response for LaAlO$_3$ and SrLaAlO$_4$ substrates agree with ARPES and Hall measurements. Our results identify Jahn--Teller distortion as a key tuning parameter in strained La$_3$Ni$_2$O$_7$ and support its central role in optimizing superconductivity in bilayer nickelates.
\end{abstract}

\maketitle
\begin{figure} 
	\begin{center}
		\includegraphics[width=\columnwidth]{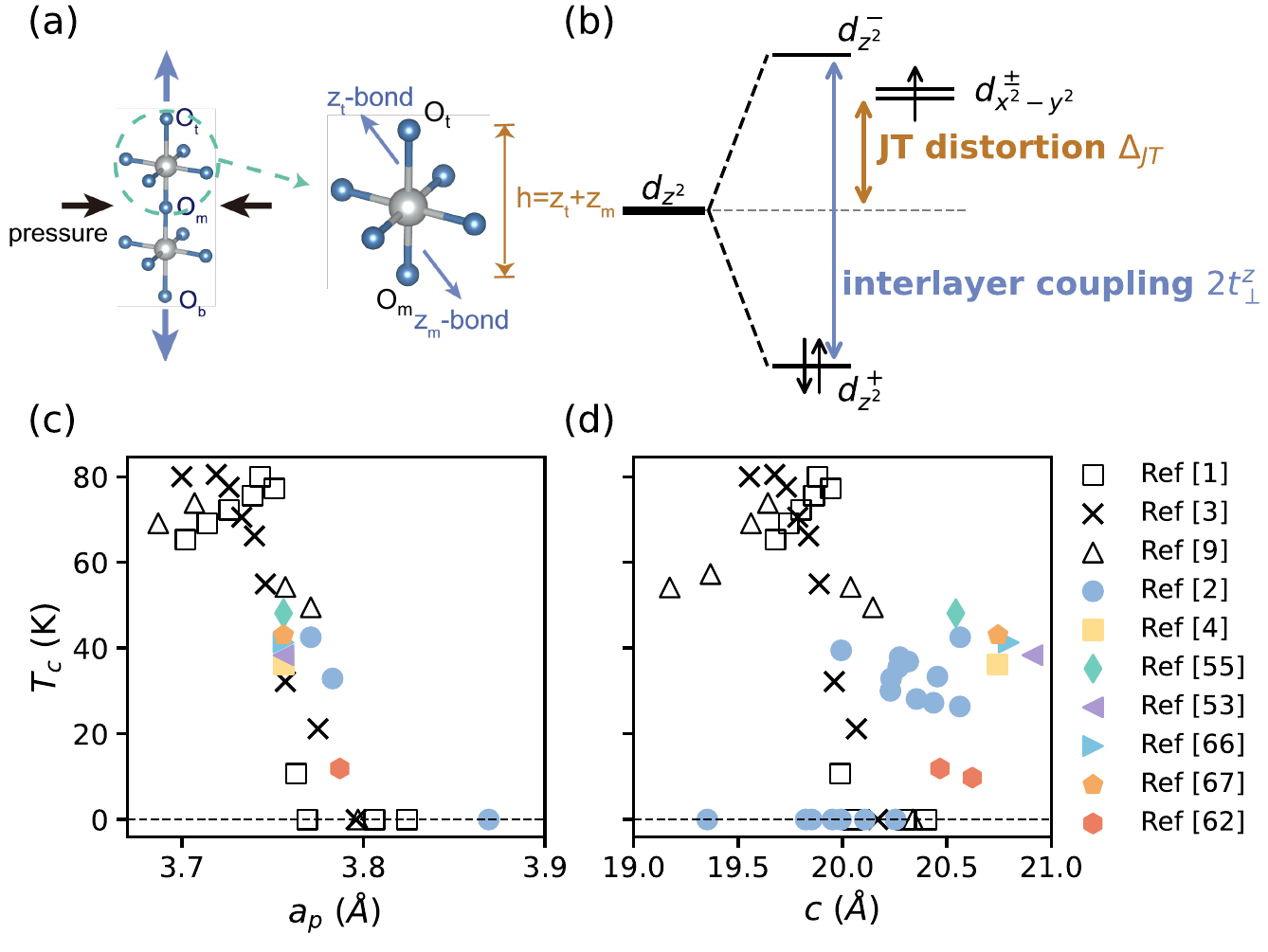}
		\caption{(a) The Ni-O octahedron elongates in the out-of-plane direction when a thin-film material is under in-plane compressive strain effect. $z_t$ represents the bond length between the Ni atom and the outer-layer (top or bottom) oxygen atom O$_{t/b}$, while $z_m$ denotes the bond length between the Ni atom and the middle oxygen atom O$_m$. The height of the NiO$_6$ octahedron is $h=z_t+z_m$. (b) Orbital energy level diagram. The JT distortion $\Delta_{JT}$ is defined as the on-site energy difference between the two $e_g$ orbitals of the Ni atom. The interlayer coupling is characterized by $t_{\perp}^{z}$. (c) Relationship between $a_{p}$ and $T_c$ from experimental data ($a_p=\sqrt{a^2+b^2}/2$ is an alternative convention used in the literature). (d) Relationship between $c$ and $T_c$ from experimental data. In (c,d), open black symbols represent bulk materials, and solid colored symbols represent thin-film materials.
}
    \label{fig1}
	\end{center}
\end{figure}

The recent discovery of superconductivity in the bilayer nickelate La$_3$Ni$_2$O$_7$ under high pressure has sparked renewed interest in layered nickelates as candidates for unconventional superconductivity~\cite{2023Wangc, 2025Hwangc, chengjg_crystal, 2025Chenb, Review, 2024Xiangb}. Unlike infinite-layer nickelate and cuprate superconductors, La$_3$Ni$_2$O$_7$ belongs to the Ruddlesden--Popper $n=2$ series and features closely spaced NiO$_2$ bilayers \cite{Review}, as illustrated in Fig.~\ref{fig1}(a). Under applied pressure, superconductivity emerges with a high transition temperature~\cite{2023Wangc, 2024Yuan, chengjg_crystal, chengjg_poly, 2025Wang, 2025chenc, zhangjunjie,  2025Chenf, 2024Chenc, 2025Wangk, 2025Maoa,  2025Cuia, 2026Yao,4310-1,4310-2,4310-3}, indicating that structural and orbital degrees of freedom beyond those of single-layer systems play a central role. Identifying the essential ingredients responsible for superconductivity in this material therefore remains an outstanding open problem.

A central issue in this context concerns the role of the Ni $d_{3z^2-r^2}$ ($d_{z^2}$) orbital. In contrast to cuprates, La$_3$Ni$_2$O$_7$ exhibits pronounced multi-orbital character, with the relevant atomic energy levels summarized in Fig.~\ref{fig1}(b)~\cite{2023Wangc, Review}. In particular, the $d_{z^2}$ orbital participates in strong interlayer coupling $t_{\perp}^z$, leading to the formation of bonding ($d_{z^2}^{+}$) and antibonding ($d_{z^2}^{-}$) molecular orbitals across the bilayer. This interlayer hybridization has been widely proposed as a key factor in stabilizing superconductivity~\cite{2023Yao,
2023Wangd, 2023Zhango, 2023Yangb, 2023Eremin, 2023Si, 2024Zhangf, 2025Xianga, 2024Wua, 2024Kurokia, GangSu_PhysRevLett.132.036502, 2024Wangi, 2024Wehling, 2024Hub-dft, 2025Jiang-dft, 2025Hua-dft, 2025Wangl, 2025Hua, 2025Hug, 2025Zhange, 2025Zhangh, 2025Li, 2025Wehling, 2026Raghu, 2024Hirschfelda, 2025Botana, 2025Hue, 2025Yaof, 2025Chen, 2025Kuroki, 2025Chaloupka, 2026Kontani}. At the same time, the importance of the Jahn--Teller (JT) splitting energy $\Delta_{JT}$ between the $d_{x^2-y^2}$ and $d_{z^2}$ orbitals has often been underestimated \cite{Review}. How the competition between interlayer coupling $t_{\perp}^z$ and JT splitting $\Delta_{JT}$ shapes the electronic structure of La$_3$Ni$_2$O$_7$ remains unresolved.

The recent realization of La$_3$Ni$_2$O$_7$ thin film superconductivity under substrate-induced strain at ambient pressure provides a new platform to address this issue~\cite{2025Hwangc, 2025Chenb, 2025Nie, 2025Tsukazaki, 2025Hwang, 2026Goodge, 2025Xuea, 2025Shen, 2025Niea, 2025Xue, 2025Hwang, 2025Hwanga, 2025Hwangb, 2026Yua,  2025He, 2026Nie, 2025Wena, 2025Chenact}. Experimentally, the crystal structures associated with superconducting La$_3$Ni$_2$O$_7$ are summarized in Fig.~\ref{fig1}(c,d)~\cite{2023Wangc,2025Hwangc,2025Chenf,2025Hwangb}. It has been proposed that superconductivity appears only when the in-plane lattice constant $a_{p}$ is reduced below a critical value \cite{2025Hwangc}, as shown in Fig.~\ref{fig1}(c). In contrast, thin-film studies indicate that the out-of-plane lattice constant $c$ is not essential for the emergence of superconductivity \cite{2025Hwanga,2026Goodge,2025Hwangb}, as shown in Fig.~\ref{fig1}(d). This difference between bulk and thin films may give important clues for realizing superconductivity in bilayer nickelates.
In this work, we present a systematic study of the electronic structure of La$_3$Ni$_2$O$_7$ thin films under strain. Our results demonstrate that JT physics plays a key role in shaping the electronic structure in La$_3$Ni$_2$O$_7$.

More precisely, the essential structural motif of La$_3$Ni$_2$O$_7$ is the network of corner-sharing NiO$_6$ octahedra forming closely spaced bilayers, as illustrated in Fig.~\ref{fig1}(a). These bilayer octahedra arrange into a square lattice within the $ab$ plane, giving rise to the quasi-two-dimensional crystal structure. In a naive local picture, one may assume that each Ni ion resides in an ideal octahedral environment with $O_h$ symmetry, which splits the Ni $3d$ orbitals into twofold-degenerate $e_g$ and threefold-degenerate $t_{2g}$ manifolds. However, the bilayer geometry introduces two crucial modifications beyond this simple picture: (i) \textit{inequivalent apical oxygen ions and JT distortions}, and (ii) \textit{strong interlayer coupling between the $d_{z^2}$ orbitals}.

From a symmetry perspective, the inner apical oxygen O$_m$ located between the two NiO$_2$ layers serves as the only inversion center (or equivalently the only mirror plane normal to the $c$ axis), as indicated in Fig.~\ref{fig1}(a). This symmetry operation relates the top and bottom Ni sites within the bilayer. In contrast, the outer apical oxygen ions O$_t$ above and O$_b$ below the bilayer become inequivalent to the inner oxygen O$_m$, leading to a reduction of the local symmetry from $O_h$ to $C_{4v}$. To quantify this asymmetry, we denote by $z_t$ and $z_m$ the Ni--O bond lengths associated with the outer and inner apical oxygen ions, respectively. The total apical height along the $c$ direction is then defined as $h = z_t + z_m$, as shown in Fig.~\ref{fig1}(a). 

Importantly, the inner apical oxygen plays a dominant role in mediating the interlayer hopping between $d_{z^2}$ orbitals. As a consequence, the interlayer coupling $t_\perp^z$ is primarily controlled by the middle oxygen O$_m$, while the outer apical oxygens O$_{t/b}$ have a much weaker influence. Taking these structural ingredients into account, the atomic level scheme for the two Ni sites is summarized in Fig.~\ref{fig1}(b). In La$_3$Ni$_2$O$_7$, the nominal Ni valence is close to $3d^{7.5+}$, which leads to an almost fully occupied bonding state $d_{z^2}^{+}$. The remaining electron occupies the $d_{x^2-y^2}^{\pm}$ and antibonding $d_{z^2}^{-}$ orbitals, depending on their relative bandwidths and energy positions.

Under epitaxial strain, in-plane compression along the $a,b$ directions drives an anisotropic distortion of the NiO$_6$ octahedra, elongating them along the $c$ axis, as illustrated in Fig.~\ref{fig1}(a). We find that the Ni--O$_t$ and Ni--O$_m$ bond lengths respond differently to the applied strain, resulting in a strong sensitivity of the electronic structure to the detailed octahedral geometry. In general, elongation (compression) along the $c$ ($a,b$) direction enhances the Jahn–Teller splitting $\Delta_{JT}$, while $t_{\perp}^{z}$ remains nearly constant under strain.

\begin{figure}[htbp!]
	\begin{center}
		\includegraphics[width=\columnwidth]{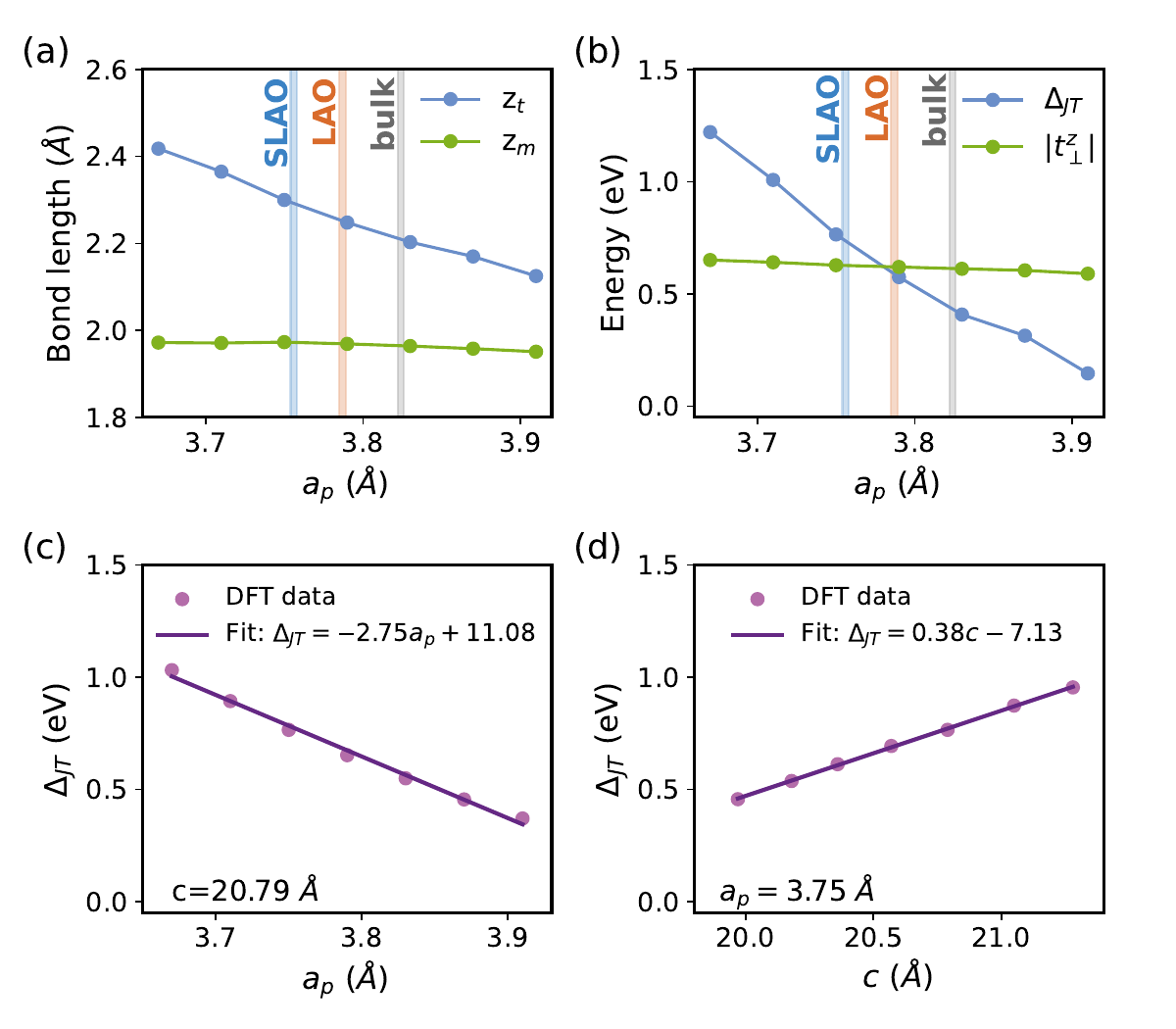}
		\caption{(a) Variation of the bond lengths $z_{t}$ (between Ni and outer-layer O$_{t}$) and $z_{m}$ (between Ni and inner-layer O$_{m}$) with respect to the in-plane lattice constant $a_{p}$. (b) Variation of $\Delta_{JT}$ and $|t_{\perp}^{z}|$ with respect to $a_{p}$. As a reference, the vertical blue, orange and gray thick line represents the experimental lattice constants associated with SLAO substrate~\cite{2025Hwang}, LAO substrate~\cite{2025Hwangb} and the bulk material under ambient pressure~\cite{2023Wangc}, respectively. (c) Variation of $\Delta_{JT}$ with the in-plane lattice parameter $a_{p}$ while keeping the lattice parameter along the $c$-axis fixed at $c=20.79\AA$. The data is fitted linearly, depicted in solid line, with the slope given by $-2.75$eV/$\AA$. (d) Variation of $\Delta_{JT}$ with the lattice parameter $c$ while keeping the in-plane lattice parameter fixed at $a_p=3.75\AA$. The data is fitted linearly, depicted in solid line, with the slope given by $0.38$eV/$\AA$.}
    \label{fig2}
	\end{center}
\end{figure}

To substantiate this observation, we perform density functional theory (DFT) calculations on strained thin films using a three-dimensional periodic crystal structure. To simplify the calculations, we adopted a structure with $I4/mmm$ symmetry, where $a_p$ equals twice the in-plane Ni–O bond length. The substrate-induced biaxial strain is simulated by fixing the in-plane lattice constant $a_{p}$, while allowing the out-of-plane lattice constant $c$ to relax. This setup effectively corresponds to an “infinitely thick” film \cite{2025Botana,2024Hirschfelda,2025Jiang-dft}. For comparison, we also consider a single-layer structure (half unit cell) \cite{2025Chen} without $c$-axis periodicity and find that our conclusions remain qualitatively unchanged (see Appendix~\ref{Appendix:monolayer} for details). For films of several nanometers thickness, where surface effects are negligible, the infinitely thick approximation is appropriate. In our calculation, the meta generalized gradient approximation (meta-GGA) exchange-correlation functional and its SCAN version \cite{sun2015strongly} is used. Compared with GGA, the exchange–correlation functional of meta-GGA depends on the second-order derivative term of the charge density, thus yielding more accurate results than GGA. We note that, for correlated electron systems, quantitatively reliable hopping parameters remain challenging to extract within the standard DFT framework. Nevertheless, DFT calculations are expected to capture the overall trends of the electronic structure. 

Focusing on the partially occupied $e_g$ orbitals from the DFT calculations, the specific values of $\Delta_{JT}$ and $t_{\perp}^{z}$ were obtained by fitting the DFT band structure using Wannier90 \cite{mostofi2008wannier90,marzari2012maximally}. Given that there are two Ni atoms per unit cell, the resulting tight-binding (TB) model is a 4-band system \cite{2023Yao,Review}. In order to keep consistent with the standard notation, its Hamiltonian, $H$, expressed in the basis  $(d_{t\textbf{k}}^{x},d_{t\textbf{k}}^{z},d_{b\textbf{k}}^{x},d_{b\textbf{k}}^{z})$ ($t$ and $b$ denote the top-layer Ni and bottom-layer Ni, respectively, and the spin index is omitted here), is written as:
\begin{align}
H & ({\textbf{k}})=\left(\begin{array}{cc}
H_{t}({\textbf{k}}) & H_{\perp}({\textbf{k}})\\
H_{\perp}({\textbf{k}}) & H_{t}({\textbf{k}})
\end{array}\right),\label{eq:tb-lp}
\end{align}
where $H_{t}({\textbf{k}})$, $H_{\perp}({\textbf{k}})$ are $2\times 2$ block matrices (more details in Appendix~\ref{Appendix:model}).
Notice that, the JT distortion term $\Delta_{JT} (d_{t\textbf{k}}^{x\dagger} d_{t\textbf{k}}^{x}-d_{t\textbf{k}}^{z\dagger} d_{t\textbf{k}}^{z})$ is contained in $H_{t}$. The interlayer term $t_\perp^z(d_{t\textbf{k}}^{z\dagger} d_{b\textbf{k}}^{z}+d_{b\textbf{k}}^{z\dagger} d_{t\textbf{k}}^{z})$ belongs to $H_{\perp}$, whereas $t_\perp^x$ between $d_{x^2-y^2}$ is relatively weak. 

Through fixing $a_{p}$ and relaxing $c$, we obtain a series of results as shown in Fig.~\ref{fig2}(a,b). The bulk value of \LNO under low pressure is $a_{p}=3.824$ \AA~\cite{2023Wangc}, which is represented by the vertical gray line. Fig.~\ref{fig2}(a) shows that upon reducing $a_{p}$, $z_{t}$ (the bond length between Ni and the outer apical oxygen O$_t$) exhibits a pronounced elongation, while $z_{m}$ (Ni--O$_{m}$ bond length for the inner apical oxygen O$_m$) remains almost intact. The contrasting behaviors in $z_t$ and $z_m$ manifest the asymmetry between the two outer apical oxygen ions and the inner oxygen ion, as discussed in previous contents.

This asymmetric structural response has direct consequences for the low-energy electronic structure, which is reflected by the extracted TB parameters, as shown in Fig.~\ref{fig2}(b). With $a_p$ decreasing and the octahedral height $h = z_t + z_m$ increasing, the energy of the $d_{x^2-y^2}$ orbital becomes higher while that of the $d_{z^2}$ orbital is lowered, leading to a strong enhancement of the Jahn--Teller splitting $\Delta_{JT}$. In contrast, the interlayer hopping $t_{\perp}^{z}$ is primarily mediated by the inner apical oxygen O$_m$ and is therefore controlled by $z_m$, which remains nearly unchanged. As a result, $t_{\perp}^{z}$ varies only weakly under strain.

A weak residual increase of $t_{\perp}^{z}$ with decreasing $a_p$, as observed from Fig.~\ref{fig2}(b), can be attributed to higher-order bonding effects involving the outer apical oxygens. To characterize this effect, we compute the integrated crystal orbital Hamilton population (ICOHP) \cite{dronskowski1993crystal,deringer2011crystal} using LOBSTER \cite{maintz2016lobster,nelson2020lobster}. The ICOHP serves as a qualitative measure of bonding strength between atomic orbitals, where more negative values indicate stronger bonding. As shown in Table~\ref{tab;tab1}, the Ni--O$_m$ bond remains substantially stronger than the Ni--O$_t$ bond throughout the strain range, confirming that O$_m$ provides the dominant hopping channel for $t_\perp^z$. Meanwhile, the Ni-O$_{t}$ bond is much more sensitive to strain, consistent with the pronounced elongation of $z_t$. Importantly, as $a_{p}$ decreases, the relative strength of the Ni-O$_{m}$ bond compared to Ni-O$_{t}$ increases, enhancing the hybridization along the inner-apical pathway and leading to a slight increase in $t_{\perp}^{z}$. However, this is a higher-order effect and does not qualitatively alter the weak strain dependence of $t_{\perp}^{z}$. These results support the picture that epitaxial strain mainly modifies the outer apical environment and hence $\Delta_{JT}$, while the inner-apical hopping channel for $t_\perp^z$ is only weakly affected.


\begin{table}[!htbp] 
\centering
\caption{ICOHP values of Ni-O$_{t}$ and Ni-O$_m$ bonds under different in-plane lattice parameters $a_{p}$. A more negative ICOHP indicates stronger bond strength.} 
\begin{ruledtabular}
\setlength{\tabcolsep}{1pt}
\begin{tabular}{cccccccc}
$a_{p}$ ($\AA$)&$3.67$&$3.71$&$3.75$&$3.79$&$3.83$&$3.87$&3.91 \\
Ni-O$_{t}$ &$-0.335$&$-0.414$&$-0.529$&$-0.636$&$-0.756$&$-0.875$&$-1.001$ \\
Ni-O$_m$ &$-1.324$&$-1.359$&$-1.388$&$-1.430$&$-1.476$&$-1.508$&$-1.573$
\end{tabular}
\end{ruledtabular}
\label{tab;tab1}
\end{table}

To further disentangle the structural origin of the JT splitting $\Delta_{JT}$, we independently vary the in-plane lattice constant $a_p$ and the out-of-plane lattice constant $c$, as shown in Fig.~\ref{fig2}(c,d). Specifically, in Fig.~\ref{fig2}(c), we fix $c=20.79$ $\AA$ (the relaxed value from DFT when $a_{p}=3.75\AA$) and change only $a_{p}$, whereas in Fig.~\ref{fig2}(d), we fix $a_p=3.75$ $\AA$ and change only $c$. We find that $\Delta_{JT}$ depends strongly on $a_p$, with a linear slope of approximately $-2.75$ eV/\AA, while its dependence on $c$ is much weaker, with a linear slope of only $0.38$ eV/\AA. This clear hierarchy demonstrates that $\Delta_{JT}$ is predominantly controlled by the in-plane lattice constant, rather than the $c$-axis distortion.

This result is directly relevant to experiments. Superconductivity in La$_2$PrNi$_2$O$_7$ thin films has been realized on SLAO ($T_{c,0}\!\approx\!30$~K) and LAO ($T_{c,0}\!\approx\!3$~K) substrates \cite{2025Hwangb,2025Hwang}. Previous studies have suggested that the LAO lattice constant lies near the critical strain condition for superconductivity \cite{2025Hwangb}, as summarized in Fig.~\ref{fig1}(c,d). Remarkably, this relatively small lattice change is accompanied by an order-of-magnitude variation in $T_{c}$. Combined with the strong $a_p$-dependence of $\Delta_{JT}$ obtained above, this suggests that the strain-induced evolution of the JT splitting provides a natural microscopic link between lattice strain and superconductivity. We note that it is possible that the critical in-plane lattice constant corresponds to a boundary of a structural transition. However, the LAO data point~\cite{2025Hwangb} appears to share similar structural characteristics with nearby samples, suggesting that structural changes alone are insufficient to account for the sharp variation in $T_c$. This indicates that both structural effects and the associated evolution of the crystal field may be relevant.

\begin{figure}
    \centering
    \includegraphics[width=\linewidth]{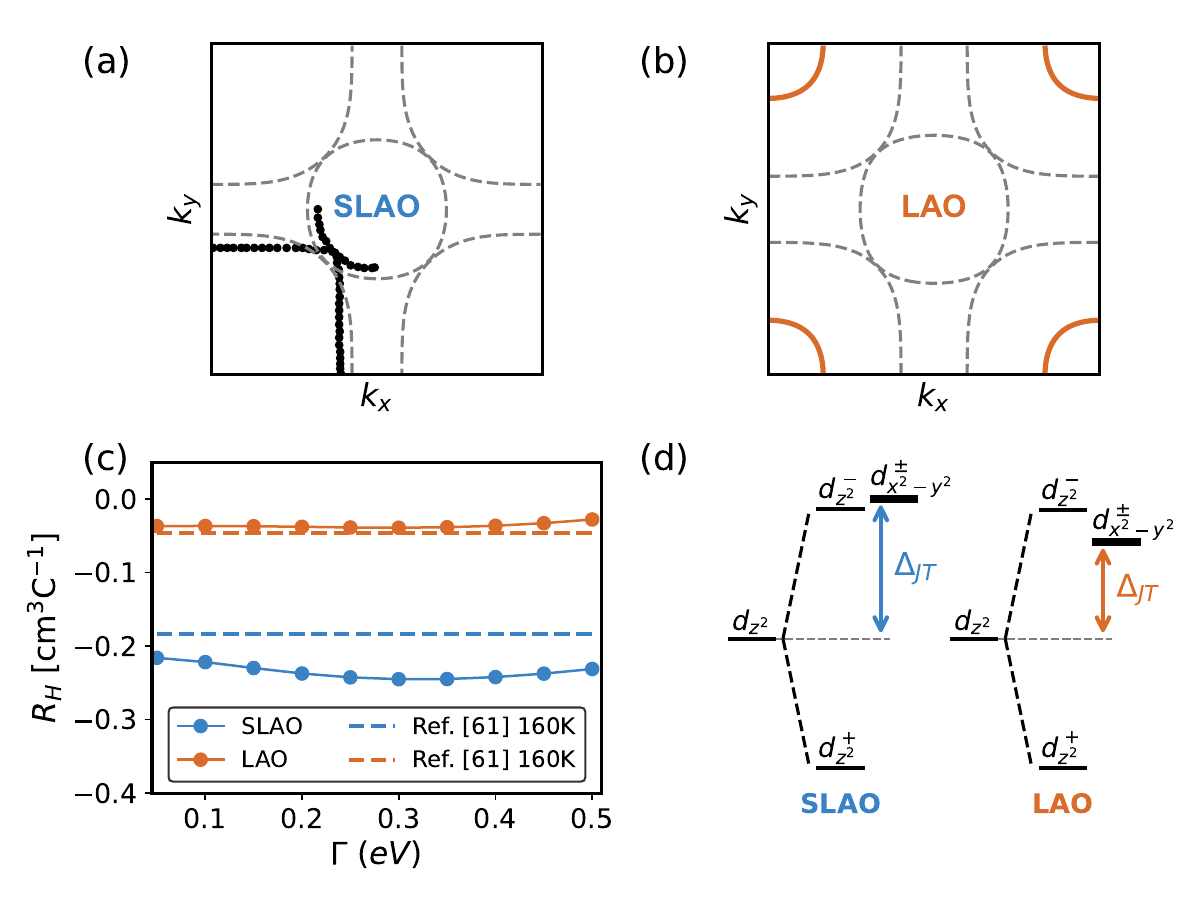}
    \caption{(a) and (b) show Fermi surface for SLAO and LAO substrate, respectively. The Fermi surface extracted from ARPES on SLAO \cite{2025Shen} is plotted as black dots in (a). The main contrast between the two substrates is the emergence of $\gamma$ pocket highlighted in LAO. (c) The Hall coefficient obtained from the two substrates as a function of quasiparticle scattering rate $\Gamma$. The dashed lines show the Hall coefficient from experimental measurements at 160K~\cite{2025Hwangb}. (d) Schematic illustration of the local electronic configuration for SLAO and LAO, where the JT distortion $\Delta_{JT}$ is labeled by arrows.}
    \label{fig:Hall}
\end{figure}

To further elucidate how the JT distortion impacts the electronic structure, we compare the band structures of thin films on SLAO and LAO substrates. The most direct consequence of the JT distortion is a qualitative reconstruction of the Fermi surface, as shown in Fig.~\ref{fig:Hall}(a,b), with TB parameters specified in Appendix~\ref{Appendix:model}. For SLAO, we obtain two Fermi surfaces: an electron-like $\alpha$-pocket and a hole-like $\beta$-pocket, whose shapes agree well with recent ARPES measurements~\cite{2025Shen}. In contrast, for LAO, an additional hole-like $\gamma$-pocket appears near the Brillouin zone corner, as highlighted in Fig.~\ref{fig:Hall}(b). As the JT distortion increases, the onsite energy of $d_{z^2}$ orbitals is lowered with respect to the $d_{x^2-y^2}$ orbital, as shown in Fig.~\ref{fig:Hall}(d). Consequently, the $\gamma$-band, which is predominantly of symmetric $d^+_{z^2}$ character near the zone corner, is pushed below the Fermi level, resulting in the disappearance of the $\gamma$-pocket. This establishes a $\Delta_{JT}$-driven Lifshitz transition associated with the appearance or disappearance of the $\gamma$-pocket.

The Fermi surface reconstruction has immediate consequences in the Hall coefficient. In particular, the disappearance of the hole-like $\gamma$ pocket and the changes in Fermi surface shapes reduce carrier compensation, leading to a more electron-dominated response. Using the obtained band structures and a uniform quasiparticle scattering rate $\Gamma$, we calculate the Hall coefficient $R_H=\sigma_{xy}/\sigma_{xx}\sigma_{yy}$~\cite{1992John} (see Appendix~\ref{Appendix: Hall} for details), with results shown as solid lines in Fig.~\ref{fig:Hall}(c). For both substrates, we find $R_H$ is negative, but significantly more negative for SLAO. This reflects the JT-controlled redistribution of carrier contributions: as $\Delta_{JT}$ increases, the $\gamma$-pocket disappears and the $\beta$-pocket contribution is reduced, while the electron-like $\alpha$-pocket remains largely unchanged. The calculated results show qualitative agreement with experimental data \cite{2025Hwangb}, plotted in dashed lines, where the magnitude of $R_H$ for LAO is about 4-5 times smaller than that of SLAO. This further shows that $\Delta_{JT}$ controls the balance between different charge carriers through shaping the band structures.

We now contrast the strained thin-film case with bulk La$_3$Ni$_2$O$_7$ under hydrostatic pressure. Unlike epitaxial strain, which primarily constrains the in-plane lattice constants, hydrostatic pressure compresses the lattice along all three directions simultaneously. The bulk case therefore provides a useful reference for assessing whether $\Delta_{JT}$ and $t_{\perp}^{z}$ can still be tuned independently.

\begin{figure} 
	\begin{center}
		\includegraphics[width=\columnwidth]{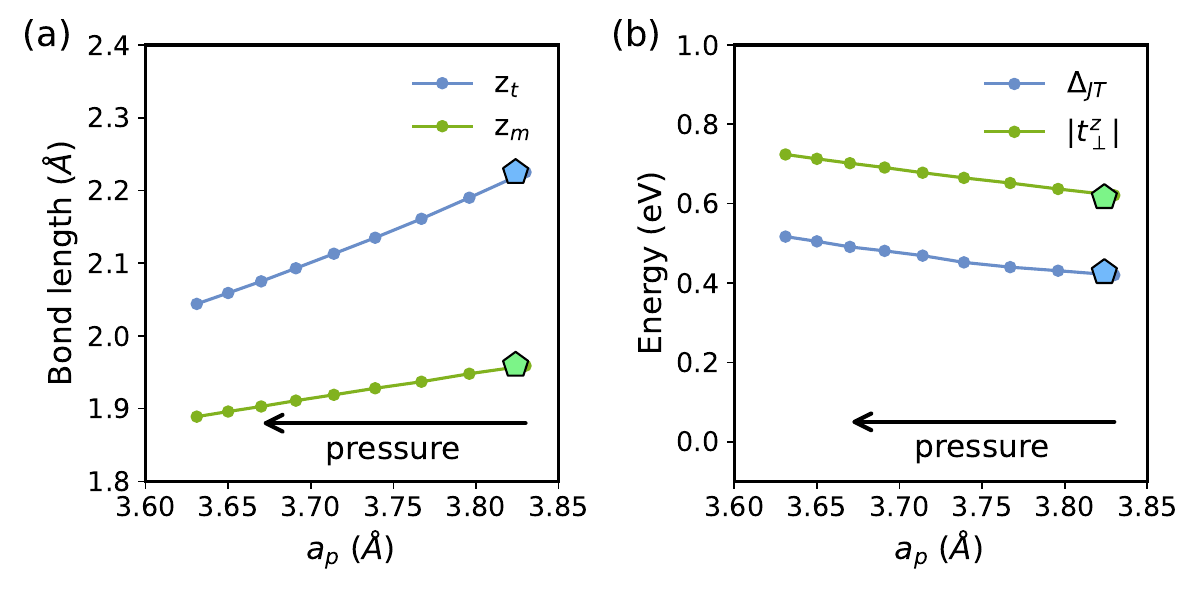}
		\caption{(a) Variation of the bond lengths $z_{t}$ (between Ni and outer-layer O$_{t}$) and $z_{m}$ (between Ni and inner-layer O$_{m}$) with the in-plane lattice constant $a_{p}$ for the bulk materials under hydrostatic pressure. (b) Variation of $\Delta_{JT}$ and $|t_{\perp}^{z}|$ with the in-plane lattice parameter $a_{p}$ in bulk case. The regular pentagon represents the values calculated using the experimental lattice constants under ambient pressure ~\cite{2023Wangc}. The black arrows indicate larger pressure.}
    \label{fig4}
	\end{center}
\end{figure}

As shown in Fig.~\ref{fig4}(a), decreasing $a_p$ in the bulk system, corresponding to increasing hydrostatic pressure, leads to a nearly uniform compression of the NiO$_6$ octahedron: both the outer apical bond length $z_t$ and the inner apical bond length $z_m$ decrease. This behavior is qualitatively different from the thin-film case, where $z_t$ increases strongly while $z_m$ remains nearly unchanged. As a consequence, the pressure dependence of $\Delta_{JT}$ and $|t_{\perp}^{z}|$ in bulk becomes much less selective, as shown in Fig.~\ref{fig4}(b). Since the in-plane lattice constant $a_p$ and the octahedral height are reduced simultaneously, the enhancement of $\Delta_{JT}$ is weakened. At the same time, the shortening of the Ni--O$_m$ bond strengthens the interlayer hybridization and increases $|t_{\perp}^{z}|$. For bulk materials under hydrostatic pressure, the changes in $\Delta_{JT}$ and $|t_{\perp}^{z}|$ are therefore of comparable magnitude.

This comparison highlights an important distinction between the two tuning routes. In thin films, epitaxial strain selectively enhances $\Delta_{JT}$ while leaving $t_{\perp}^{z}$ nearly unchanged, thereby isolating the role of JT physics in the low-energy electronic structure. In bulk materials under pressure, by contrast, both parameters evolve simultaneously, making it more difficult to distinguish their roles. The thin-film platform therefore provides a cleaner route to identify the importance of JT distortion in bilayer nickelates.

In summary, our calculations and analyses show that epitaxial strain in \LNO thin films primarily enhances the JT splitting $\Delta_{JT}$, while the interlayer hopping $t_\perp^z$ changes only weakly. This behavior originates from the asymmetric response of the apical Ni--O bonds: the outer bond is strongly elongated by in-plane compression, whereas the inner bond remains nearly unchanged. The resulting increase of $\Delta_{JT}$ controls the low-energy electronic structure and accounts for the substrate dependence of the Fermi surface and Hall response. In particular, the calculated results for LAO and SLAO are consistent with ARPES and Hall measurements. These results identify JT distortion as the leading electronic effect of epitaxial strain and a key tuning parameter for superconductivity in strained \LNO and related bilayer nickelates.

\textit{Acknowledgement.}
We acknowledge the support by the National Natural Science Foundation of China (Grant NSFC-12494594, and NSFC-12574150), the Chinese Academy of Sciences Project for Young Scientists in Basic Research (2022YSBR-048), the Innovation program for Quantum Science and Technology (Grant No. 2021ZD0302500), and Chinese Academy of Sciences under contract No. JZHKYPT-2021-08.

\bibliographystyle{apsrev4-2}
\bibliography{reference}

\newpage

\appendix
\setcounter{secnumdepth}{1}
\setcounter{figure}{0}
\renewcommand{\thefigure}{A\arabic{figure}}
\setcounter{table}{0}
\renewcommand{\thetable}{A\arabic{table}}

\section{DFT details}
Our DFT calculations employ the Vienna ab-initio simulation package (VASP) code \cite{kresse1996efficient} with the projector augmented wave (PAW) method \cite{kresse1999ultrasoft}. The cutoff energy for expanding the wave functions into a plane-wave basis is set to be 500 eV. The energy convergence criterion is 10$^{−8}$ eV. The $\Gamma$-centered 13$\times$13$\times$13 \textbf{k}-meshes are used for crystal structure with three-dimensional periodicity and 13$\times$13$\times$1 \textbf{k}-meshes are used for monolayer. 

\begin{figure}  
	\begin{center}
		\includegraphics[width=\columnwidth]{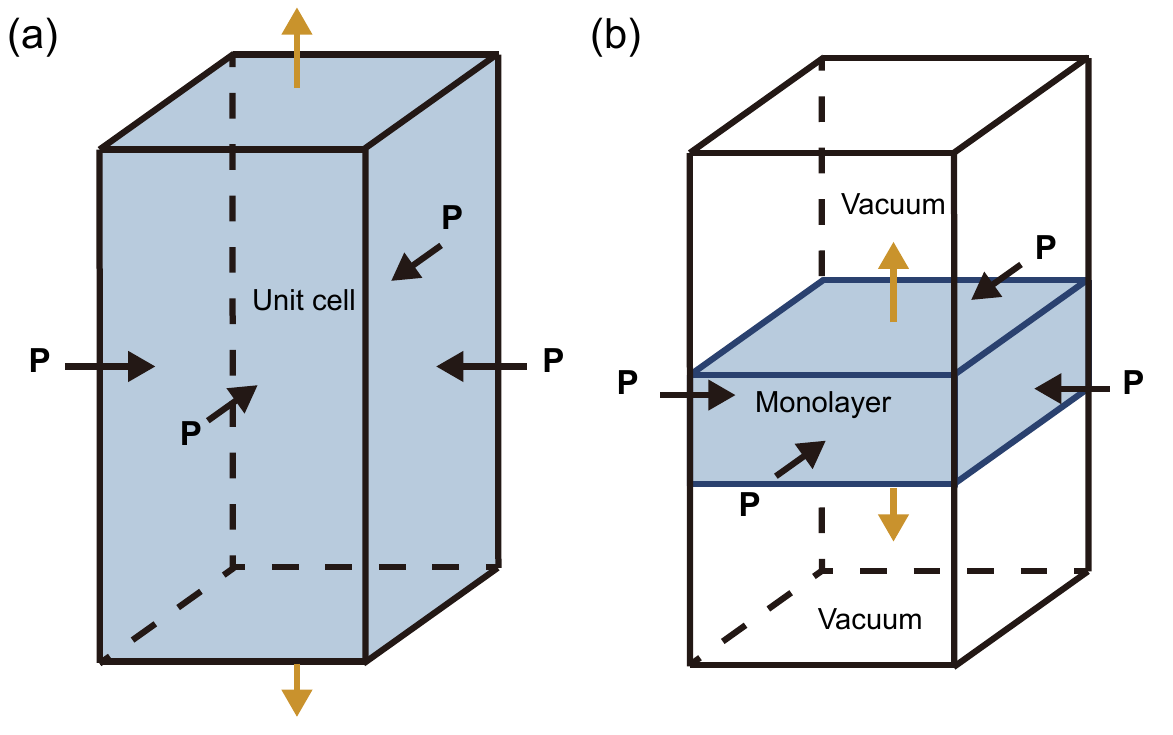}
		\caption{Two different effective methods for simulating thin-film systems. (a) Illustration of the simulation method for the "infinitely thick" film. (b) Illustration of the simulation method for the monolayer. The black border represents the unit cell used in the simulation. The blue rectangular cuboid represents the material within the unit cell. The black arrows represent the stress applied by this computational method.}
    \label{figS0}
	\end{center}
\end{figure}

The two methods for simulating thin films are shown in Fig.~\ref{figS0}. In the main text, we adopt the computational method illustrated in Fig.~\ref{figS0}(a), using a crystal structure with three-dimensional periodicity but fixing $a$ and $b$ while allowing $c$ to relax freely, which corresponds to applying a biaxial strain. In Appendix \ref{Appendix:monolayer}, we employ the computational method shown in Fig.~\ref{figS0}(b). In this method, all lattice parameters are fixed in the software while the internal atoms are allowed to relax freely. Due to the presence of a vacuum layer, this method also effectively applies a biaxial strain to the monolayer material.

\section{Tight-binding model}\label{Appendix:model}
In the main text, the TB model has the following form:
\begin{align}
H & ({\textbf{k}})=\left(\begin{array}{cc}
H_{t}({\textbf{k}}) & H_{\perp}({\textbf{k}})\\
H_{\perp}({\textbf{k}}) & H_{t}({\textbf{k}})
\end{array}\right),\label{eq:tb-lp}
\end{align}
where $H_{t}({\textbf{k}})$, $H_{\perp}({\textbf{k}})$ are $2\times 2$ block matrices. They take the structures of
\begin{align}
H_{t}({\textbf{k}})=\left(\begin{array}{cccc}
T_{{\textbf{k}}}^{x} & V_{{\textbf{k}}}\\
V_{{\textbf{k}}} & T_{{\textbf{k}}}^{z}
\end{array}\right), \label{ht}
\end{align}
and
\begin{align}
H_{\perp}({\textbf{k}})=\left(\begin{array}{cc}
t_{\bot}^{x} & V_{{\textbf{k}}}^{\prime}\\
V_{{\textbf{k}}}^{\prime} & t_{\bot}^{z}
\end{array}\right). 
\end{align}
Here,
$T_{{\textbf{k}}}^{x/z}=t_{1}^{x/z}\gamma_k+t_{2}^{x/z}\alpha_k+\epsilon^{x/z}$, $V_{\textbf{k}}=t_{3}^{xz}\beta_k$, $V_{\textbf{k}}^{\prime}=t_{4}^{xz}\beta_k$ with $\gamma_k=2(\cos k_x+\cos k_y)$, $\alpha_k=4\cos k_x\cos k_y$, $\beta_k=2(\cos k_x-\cos k_y)$. And $\Delta_{JT}=\epsilon^{x}-\epsilon^{z}$. The parameters used for SLAO substrate and LAO substrate in Fig.~\ref{fig:Hall} of the main text are listed in Table~\ref{TB}. In the DFT calculations from which this parameter was obtained, we considered 1/3 Pr doping as in the real experiment using the virtual crystal approximation (VCA) method. The use of this method leads to the value of $\Delta_{JT}$ being slightly lower than in the case without doping. However, this does not affect the comparison here.

\begin{table*}  
    \begin{tabular}{c|cccccccccc}
		\hline \hline 
		hopping parameters (eV) & $t_{1}^{x}$ & $t_{1}^{z}$ & $t_{2}^{x}$ & $t_{2}^{z}$ & $t_{3}^{xz}$ & $t_{4}^{xz}$ & $t_{\perp}^{x}$ & $t_{\perp}^{z}$ & $\epsilon^{x}$ & $\epsilon^{z}$ \\
        \hline
        SLAO & $-0.4732$ & $-0.0759$ & $0.0779$ & $-0.0154$ & $0.2022$ & $-0.0263$ & $0.0081$ & $-0.5868$ & $0.9416$ & $0.3072$ \\ 
        \hline
        LAO & $-0.4609$ & $-0.0877$ & $0.0743$ & $-0.0158$ & $0.2106$ & $-0.0294$ & $0.0087$ & $-0.5845$ & $0.8011$ & $0.3631$ \\
        \hline
        \hline 
	\end{tabular}

 \caption{The hopping parameters of the TB Hamiltonian corresponding to the \LNO thin film on SLAO substrate and LAO substrate used in Fig.~\ref{fig:Hall} of the main text.}
 \label{TB}
\end{table*}

The main difference between the two is $\Delta_{JT}$: $\Delta_{JT}$ = 0.6344 eV for the SLAO substrate and $\Delta_{JT}$ = 0.438 eV for the LAO substrate. Other than that, the remaining parameters are nearly unchanged.

\begin{figure}  
	\begin{center}
		\includegraphics[width=\columnwidth]{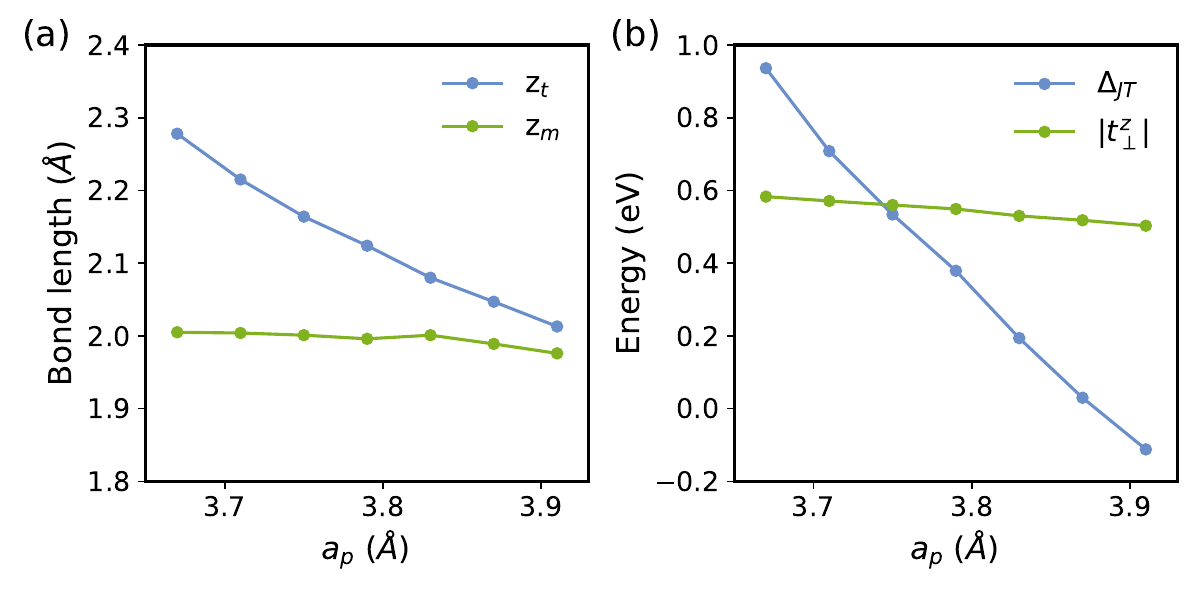}
		\caption{(a) Variation of the bond lengths between Ni and outer-layer O$_{t}$ ($z_{t}$) and between Ni and inner-layer O$_{m}$ ($z_{m}$) with the in-plane lattice constant $a_{p}$ for the monolayer crystal structure under in-plane pressure. (b) Variation of $\Delta_{JT}$ and $|t_{\perp}^{z}|$ with the in-plane lattice parameter $a_{p}$ in monolayer case. }
    \label{figS1}
	\end{center}
\end{figure}

\section{The results of monolayer}\label{Appendix:monolayer}
We also perform calculations in the opposite limit—the monolayer case—and present the results in Fig.~\ref{figS1}. As shown, the trends of the physical quantities are qualitatively consistent with those in the main text: when pressure is applied, $z_t$ changes while $z_m$ remains essentially unchanged, resulting in a change in $\Delta_{JT}$ while $t_{\perp}^{z}$ remains essentially unchanged. The remaining quantitative differences can be attributed to enhanced surface effects in the single-layer geometry. For the monolayer case, the values of $\Delta_{JT}$ and $|t_{\perp}^{z}|$ are slightly smaller than those in the main text. Taken together, these results suggest that realistic thin films with finite thickness should exhibit the same overall behavior.


\section{Hall Coefficient $R_H$}\label{Appendix: Hall}
The Hall coefficient is calculated as:
\begin{equation}
    R_H = \frac{\sigma_{xy}}{\sigma_{xx}\sigma_{yy}},
\end{equation}
where $\sigma_{xy}$ is the Hall conductivity and $\sigma_{xx,yy}$ denotes the longitudinal conductivity. We take $e$ as the electron charge, $V$ the number of the $k$-points in the Brillouin zone, $\epsilon_\bp$ the dispersion obtained from DFT, $\mu$ the chemical potential and $\Gamma$ the quasiparticle scattering rate. The expression for longitudinal conductivity reads:
\begin{equation}
    \sigma_{\alpha\alpha}=-\frac{e^2}{V}\sum_\bp (\epsilon_\bp^\alpha)^2 \lim_{\omega\rightarrow0} \text{Im}[\Pi(\bp,\omega)/\omega].
\end{equation}
Here
\begin{equation}
    \lim_{\omega\rightarrow0}\text{Im}[\Pi(\bp,\omega)/\omega]= \int_{-\infty}^{\infty}\frac{d\epsilon}{\pi} \frac{\partial f}{\partial \epsilon} \frac{\Gamma^2}{[(\epsilon - \epsilon_\bp+\mu)^2 + \Gamma^2]^2}.
\end{equation}
The expression for Hall conductivity reads:
\begin{equation}
    \sigma_H^{xyz}=\frac{e^3}{2V}\sum_\bp \epsilon_\bp^x\varepsilon^{z\gamma\delta}\epsilon_\bp^\gamma \epsilon_{\bp}^{y\delta} \lim_{\omega\rightarrow0}\Pi_H(\bp,\omega)/\omega,
\end{equation}
where
\begin{equation}
    \lim_{\omega\rightarrow0}\Pi_H(\bp,\omega)/\omega = \frac43\int_{-\infty}^{\infty}\frac{d\epsilon}{\pi} \frac{\partial f}{\partial \epsilon} \frac{\Gamma^3}{[(\epsilon - \epsilon_\bp+\mu)^2 + \Gamma^2]^3}.
\end{equation}

\end{document}